\documentclass[12pt,a4paper]{article}

\usepackage{latexsym}
\usepackage{amsmath}
\usepackage{amsfonts}
\usepackage{amssymb}
\usepackage{amsthm}
\usepackage{amscd}
\usepackage{mathrsfs}

\def\si{\sigma}                         %
\def\lm{\lambda}                        %
\def\bC{{\mathbb C}}                    %
\def\beq{\begin{equation}}              %
\def\eeq{\end{equation}}                %
\begin{document}

\newtheorem{theorem}{Theorem}[section]

\vspace*{0.5cm}
\begin{center}
{\Large \bf A note on the Gauss decomposition  of the elliptic  Cauchy   matrix}

\end{center}

\vspace{0.2cm}

\begin{center}
L. Feh\'er${}^{a}$, C. Klim\v c\'\i k${}^b$ and S. Ruijsenaars${}^{c}$ \\

\bigskip

${}^a$Department of Theoretical Physics, MTA  KFKI RMKI,\\
1525 Budapest 114, P.O.B. 49,  Hungary, and\\
Department of Theoretical Physics, University of Szeged,\\
Tisza Lajos krt 84-86, H-6720 Szeged, Hungary\\
e-mail: lfeher@rmki.kfki.hu

${}^b$ Institute de math\'ematiques de Luminy,
 \\ 163, Avenue de Luminy, \\ 13288 Marseille, France\\
 e-mail: ctirad.klimcik@univmed.fr

 ${}^{c}$Department of Applied Mathematics, University of Leeds,\\ Leeds LS2 9JT,  UK\\
e-mail: siru@maths.leeds.ac.uk

\bigskip

\end{center}

\vspace{0.2cm}

\begin{abstract} Explicit formulas for the  Gauss decomposition of elliptic Cauchy type matrices
are derived in a very simple way.
The  elliptic Cauchy identity is an immediate  corollary.

\end{abstract}

\newpage

\section{Introduction}

 The matrix
 \beq\label{Caum}
 C=\left( \frac{1}{q_i-r_j}\right)_{i,j=1}^N,\ \ \ \ q,r\in\bC^N,
 \eeq
has determinant
\beq\label{Cau}
|C|= \frac{\prod_{1\le i<j\le N}(q_i-q_j)(r_j-r_i)}{\prod_{1\le i,j\le N}(q_i-r_j)}.
\eeq
This identity was first obtained by Cauchy. Now well-known as Cauchy's identity,
it has found applications in harmonic analysis, soliton theory, and relativistic Calogero-Moser systems.

Its elliptic generalization involving Weierstrass' sigma function $\sigma(z)$ is less widely known.
It is given by
\beq\label{Frob}
\det\left(\frac{\si(q_i-r_j+\lm)}{\si(\lm)\si(q_i-r_j)}\right)_{i,j=1}^N=\frac{\si(\lm +\sum_{k=1}^N(q_k-r_k))}{\si(\lm)}
\frac{\prod_{1\le i<j\le N} \si(q_i-q_j)\si(r_j-r_i)}{\prod_{1\le i,j\le N}\si(q_i-r_j)}.
\eeq
This elliptic Cauchy identity dates back to a paper by Frobenius~\cite{Fro}.
Like~\eqref{Cau}, it has shown up in various contexts, giving rise to different proofs, cf.~Refs.~\cite{Fay,Rui,Rai,KaNo}.

Clearly, the identity applies to any minor as well. Moreover, after multiplication from the left and right by
diagonal matrices (leading to so-called Cauchy-like matrices) one can still evaluate minors explicitly.

Our perspective,
 which eventually led to this note, stems from the study of
the  Lax matrices of   Calogero-Moser type systems.  In particular, we wished   to find the Gauss decomposition of
 the elliptic Cauchy-like matrix $C_N(\lambda)$ given by~\eqref{CNl} below, i.~e., to represent it as
\beq\label{Cdec}
C_N(\lambda)=UDL,
\eeq
where $U$, $D$ and $L$ are upper-triangular, diagonal and lower-triangular matrices, respectively.

In principle, this decomposition can be obtained by invoking two previously known results. Specifically, the Frobenius
formula~\eqref{Frob} can be combined with a theorem saying
that the  elements of the relevant upper- and  lower-triangular matrices can be expressed in terms of
appropriate minors of the matrix to be decomposed \cite{Y}. Indeed, as already mentioned,
 the minors of an elliptic Cauchy-like matrix also follow from the  Frobenius formula.

 In this note, we wish to report an alternative method to decompose
$C_N(\lm)$ which we
  find interesting and insightful. First of all, it is very economic,
  inasmuch as an exposition of the proof of the general
 Gauss decomposition formula and whichever of the known proofs of the Frobenius formula would require far more space and
 time. Secondly, our direct Gauss decomposition  leads to a remarkably simple new proof of the Frobenius formula itself,
 and also reproduces some other results of interest as easy consequences.

\section{The decomposition formula}

Our proof of the  following decomposition formula is self-contained, except for its use of the 3-term
identity of the $\sigma$-function,
\beq\label{tt}
\si(z+a)\si(z-a)\si(b+c)\si(b-c)+\si(z+b)\si(z-b)\si(c+a)\si(c-a)+\si(z+c)\si(z-c)\si(a+b)\si(a-b)=0.
\eeq
We recall that this identity follows directly from the well-known relation between the
Weierstrass $\wp$-function and the $\si$-function,
\beq\label{wpsi}
\wp(x)-\wp(y)=\frac{\si(y+x)\si(y-x)}{\si(x)^2\si(y)^2},
\eeq
cf.~\cite{WW}. (Indeed, one need only divide~\eqref{tt} by $\si(z)^2\si(a)^2\si(b)^2\si(c)^2$ and use~\eqref{wpsi}.)

\medskip\noindent
{\bf Theorem.}
\emph{Let $q_1,...,q_N,r_1,...,r_N,\lambda$ be complex
variables and introduce
\beq\label{lmk}
\lambda_N=\lm,\quad \lm_{k-1} := \lambda_{k} + q_k - r_k \equiv
\lm +
\sum_{j=k}^N (q_j-r_j),\ \ \ \ k=1,\ldots, N.
\eeq
 Define the elliptic Cauchy-like matrix  $C_N(\lm)$  by
\beq\label{CNl}
C_N^{ij}(\lm):=\biggl(\prod_{k=i+1}^N\frac{\si(q_i-r_k)}{\si(q_i-q_k)}\biggr)\frac{\si(q_i-r_j+\lm)}{\si(\lm)\si(q_i-r_j)}
\biggl(\prod_{l=j+1}^N \frac{\si(q_l-r_j)}{\si(r_l-r_j)}\biggr),
\quad i,j=1,...,N, \eeq where is is understood that $\prod_{k=N+1}^N
...\equiv 1$.
 Then the decomposition~\eqref{Cdec} is given by
\beq\label{UDL}
 (D^{ii})^{-1}=U^{ii}=L^{ii} =C_i^{ii}(\lm_i); \quad U^{ij}=C_j^{ij}(\lm_j),\quad  i<j; \qquad L^{ij}=C_i^{ij}(\lm_i),\quad  i>j.
 \eeq
}\begin{proof} Substituting
\beq
2z=q_i-r_j+q_N-r_N, 2a= q_i+r_j-q_N-r_N,  2b= q_i-r_j+q_N-r_N+2\lm,
2c=q_i-r_j-q_N+r_N
\eeq
 in~\eqref{tt}, we obtain an identity from which the first step
\beq\label{Cprod}
C_N(\lambda_N) = \left(\begin{matrix} I_{N-1}&c_{N}(\lm_N)\cr 0&{C_N^{NN}(\lm_N)}
\end{matrix}\right)\left(\begin{matrix} C_{N-1}(\lm_{N-1})&0\cr 0&\frac{1}{C_N^{NN} (\lm_N)}
\end{matrix}\right) \left(\begin{matrix}  I_{N-1}&0\cr  \gamma_{N}(\lm_N) &C_N^{NN}(\lm_N)\end{matrix}\right)
\eeq of an inductive decomposition follows by a straightforward
computation.
 Here, $I_{N-1}$ stands for the unit $(N-1)\times (N-1)$ matrix,  $c_{N}$ is
a column  vector whose $(N-1)$
components are $C_N^{iN}$, $i=1,...,N-1$,
 and
$\gamma_{N}$ is a row  vector whose $(N-1)$ components are $C_N^{Nj}$, $j=1,...,N-1$.

Applying the  decomposition~\eqref{Cprod} to the Cauchy matrix $C_{N-1}(\lm_{N-1})$ and
then to the Cauchy  matrix $C_{N-2}(\lm_{N-2})$ etc.,  we arrive directly
at the   formula~\eqref{Cdec} with factors~\eqref{UDL}.
 \end{proof}

\medskip

It remains to discuss some consequences.
First of all, the  following result follows effortlessly.

\medskip
\noindent {\bf Corollary (elliptic Cauchy identity).}  \emph{The formulas  \eqref{Cdec}  and \eqref{UDL}  imply
\beq
 \det (C_N(\lm)) =\prod_{k=1}^NC_k^{kk}(\lm_k)=\prod_{k=1}^N\frac{\si(\lm_{k-1})}{\si(\lm_k)\si(q_k-r_k)}=
 \frac{\si(\lm+\sum_{k=1}^N(q_k-r_k))}{\si(\lm)\prod_{k=1}^N\si(q_k-r_k)},
 \eeq
 which is just the identity \eqref{Frob} on account of \eqref{CNl} .}
\medskip

Secondly, it is worth noting that the decomposition formula provided by the Theorem  remains valid
if we replace the $\sigma$-function by
any non-zero odd
holomorphic function that satisfies the 3-term identity \eqref{tt}.
(Indeed, our proof only uses these properties of $\si(z)$.)
It is known \cite{WW}
that all such functions are of the form
\beq
\tilde \sigma(z) = e^{\alpha+ \beta z^2}\sigma(z),
\eeq
where $\alpha$ and $\beta$
are arbitrary   complex numbers, and where it is understood that the rational and trigonometric/hyperbolic
degenerations of the $\si$-function are included.
If we replace $\sigma(z)$ by the rational degeneration furnished by
$\tilde \sigma(z) = z$ and
take $\lambda$ to infinity,  then we
obtain the Gauss decomposition of the original Cauchy matrix \eqref{Caum} as well as the determinant formula \eqref{Cau}
from our result.

Thirdly, from the trigonometric  specialisation
we can recover the Gauss decomposition of the Lax matrix of the relativistic trigonometric Calogero-Moser system
that recently cropped up in the paper \cite{FK} written by two of us.
In fact, it was our discussion of the decomposition of the latter matrix that eventually led to the
general decomposition encoded in the above Theorem.

Finally, we point out that a similarity transformation with the reversal permutation  matrix can
be applied to \eqref{Cdec} to obtain a `lower-diagonal-upper' version of the decomposition formula. After a relabeling
\beq
p_1,\ldots,p_N\to p_N,\ldots,p_1,\ \ \ \ \  p=q,r,
\eeq
 this decomposition has a well-defined limit for  $N\to \infty$, by contrast to the one in the Theorem.

\bigskip
\bigskip
\bigskip
\medskip
\noindent{\bf Acknowledgements.}
This work was supported in part
by the Hungarian
Scientific Research Fund (OTKA) under the grant K 77400.

\end{document}